\RequirePackage{lineno}
\documentclass[aps,prl,twocolumn,showpacs]{revtex4-1}
\usepackage{times}
\usepackage{amsmath}
\usepackage{graphicx}
\usepackage{color}
\usepackage{multirow}
\usepackage[T1]{fontenc}

\newcommand{\pp}{\pi^+\pi^-}

\newcommand{\kk}{K^+K^-}

\newcommand{\EE}{e^+e^-}
\newcommand{\MM}{\mu^+\mu^-}

\newcommand{\jpsi}{J/\psi}
\newcommand{\piz}{\pi^0}
\newcommand{\chiz}{\chi_{c0}}
\newcommand{\chit}{\chi_{c2}}

\usepackage{xcolor} 
\usepackage[colorlinks=true,linkcolor=blue,citecolor=blue,urlcolor=blue]{hyperref} %

\begin{document}

\title{
\boldmath Precise Measurement of the $\chiz$ Resonance Parameters and Branching Fractions of $\chi_{c0,c2}\to\pp/\kk$
}

\author{
\begin{small}
\begin{center}
M.~Ablikim$^{1}$, M.~N.~Achasov$^{4,c}$, P.~Adlarson$^{76}$, O.~Afedulidis$^{3}$, X.~C.~Ai$^{81}$, R.~Aliberti$^{35}$, A.~Amoroso$^{75A,75C}$, Y.~Bai$^{57}$, O.~Bakina$^{36}$, I.~Balossino$^{29A}$
% ...
\\
\vspace{0.2cm}
(BESIII Collaboration)\\
\vspace{0.2cm} {\it
	$^{1}$ Institute of High Energy Physics, Beijing 100049, People's Republic of China\\
	$^{2}$ Beihang University, Beijing 100191, People's Republic of China\\
	$^{3}$ Bochum  Ruhr-University, D-44780 Bochum, Germany\\
	$^{4}$ Budker Institute of Nuclear Physics SB RAS (BINP), Novosibirsk 630090, Russia\\
	$^{5}$ Carnegie Mellon University, Pittsburgh, Pennsylvania 15213, USA\\
	$^{6}$ Central China Normal University, Wuhan 430079, People's Republic of China\\
	$^{7}$ Central South University, Changsha 410083, People's Republic of China\\
	$^{8}$ China Center of Advanced Science and Technology, Beijing 100190, People's Republic of China\\
	$^{9}$ China University of Geosciences, Wuhan 430074, People's Republic of China\\
	$^{10}$ Chung-Ang University, Seoul, 06974, Republic of Korea\\
	$^{11}$ COMSATS University Islamabad, Lahore Campus, Defence Road, Off Raiwind Road, 54000 Lahore, Pakistan\\
	$^{12}$ Fudan University, Shanghai 200433, People's Republic of China\\
	$^{13}$ GSI Helmholtzcentre for Heavy Ion Research GmbH, D-64291 Darmstadt, Germany\\
	$^{14}$ Guangxi Normal University, Guilin 541004, People's Republic of China\\
	$^{15}$ Guangxi University, Nanning 530004, People's Republic of China\\
	$^{16}$ Hangzhou Normal University, Hangzhou 310036, People's Republic of China\\
	$^{17}$ Hebei University, Baoding 071002, People's Republic of China\\
	$^{18}$ Helmholtz Institute Mainz, Staudinger Weg 18, D-55099 Mainz, Germany\\
	$^{19}$ Henan Normal University, Xinxiang 453007, People's Republic of China\\
	$^{20}$ Henan University, Kaifeng 475004, People's Republic of China\\
	$^{21}$ Henan University of Science and Technology, Luoyang 471003, People's Republic of China\\
	$^{22}$ Henan University of Technology, Zhengzhou 450001, People's Republic of China\\
	$^{23}$ Huangshan College, Huangshan  245000, People's Republic of China\\
	$^{24}$ Hunan Normal University, Changsha 410081, People's Republic of China\\
	$^{25}$ Hunan University, Changsha 410082, People's Republic of China\\
	$^{26}$ Indian Institute of Technology Madras, Chennai 600036, India\\
	$^{27}$ Indiana University, Bloomington, Indiana 47405, USA\\
	$^{28}$ INFN Laboratori Nazionali di Frascati , (A)INFN Laboratori Nazionali di Frascati, I-00044, Frascati, Italy; (B)INFN Sezione di  Perugia, I-06100, Perugia, Italy; (C)University of Perugia, I-06100, Perugia, Italy\\
	$^{29}$ INFN Sezione di Ferrara, (A)INFN Sezione di Ferrara, I-44122, Ferrara, Italy; (B)University of Ferrara,  I-44122, Ferrara, Italy\\
	$^{30}$ Inner Mongolia University, Hohhot 010021, People's Republic of China\\
	$^{31}$ Institute of Modern Physics, Lanzhou 730000, People's Republic of China\\
	$^{32}$ Institute of Physics and Technology, Peace Avenue 54B, Ulaanbaatar 13330, Mongolia\\
	$^{33}$ Instituto de Alta Investigaci\'on, Universidad de Tarapac\'a, Casilla 7D, Arica 1000000, Chile\\
	$^{34}$ Jilin University, Changchun 130012, People's Republic of China\\
	$^{35}$ Johannes Gutenberg University of Mainz, Johann-Joachim-Becher-Weg 45, D-55099 Mainz, Germany\\
	$^{36}$ Joint Institute for Nuclear Research, 141980 Dubna, Moscow region, Russia\\
	$^{37}$ Justus-Liebig-Universitaet Giessen, II. Physikalisches Institut, Heinrich-Buff-Ring 16, D-35392 Giessen, Germany\\
	$^{38}$ Lanzhou University, Lanzhou 730000, People's Republic of China\\
	$^{39}$ Liaoning Normal University, Dalian 116029, People's Republic of China\\
	$^{40}$ Liaoning University, Shenyang 110036, People's Republic of China\\
	$^{41}$ Nanjing Normal University, Nanjing 210023, People's Republic of China\\
	$^{42}$ Nanjing University, Nanjing 210093, People's Republic of China\\
	$^{43}$ Nankai University, Tianjin 300071, People's Republic of China\\
	$^{44}$ National Centre for Nuclear Research, Warsaw 02-093, Poland\\
	$^{45}$ North China Electric Power University, Beijing 102206, People's Republic of China\\
	$^{46}$ Peking University, Beijing 100871, People's Republic of China\\
	$^{47}$ Qufu Normal University, Qufu 273165, People's Republic of China\\
	$^{48}$ Renmin University of China, Beijing 100872, People's Republic of China\\
	$^{49}$ Shandong Normal University, Jinan 250014, People's Republic of China\\
	$^{50}$ Shandong University, Jinan 250100, People's Republic of China\\
	$^{51}$ Shanghai Jiao Tong University, Shanghai 200240,  People's Republic of China\\
	$^{52}$ Shanxi Normal University, Linfen 041004, People's Republic of China\\
	$^{53}$ Shanxi University, Taiyuan 030006, People's Republic of China\\
	$^{54}$ Sichuan University, Chengdu 610064, People's Republic of China\\
	$^{55}$ Soochow University, Suzhou 215006, People's Republic of China\\
	$^{56}$ South China Normal University, Guangzhou 510006, People's Republic of China\\
	$^{57}$ Southeast University, Nanjing 211100, People's Republic of China\\
	$^{58}$ State Key Laboratory of Particle Detection and Electronics, Beijing 100049, Hefei 230026, People's Republic of China\\
	$^{59}$ Sun Yat-Sen University, Guangzhou 510275, People's Republic of China\\
	$^{60}$ Suranaree University of Technology, University Avenue 111, Nakhon Ratchasima 30000, Thailand\\
	$^{61}$ Tsinghua University, Beijing 100084, People's Republic of China\\
	$^{62}$ Turkish Accelerator Center Particle Factory Group, (A)Istinye University, 34010, Istanbul, Turkey; (B)Near East University, Nicosia, North Cyprus, 99138, Mersin 10, Turkey\\
	$^{63}$ University of Bristol, (A)H H Wills Physics Laboratory; (B)Tyndall Avenue; (C)Bristol; (D)BS8 1TL\\
	$^{64}$ University of Chinese Academy of Sciences, Beijing 100049, People's Republic of China\\
	$^{65}$ University of Groningen, NL-9747 AA Groningen, The Netherlands\\
	$^{66}$ University of Hawaii, Honolulu, Hawaii 96822, USA\\
	$^{67}$ University of Jinan, Jinan 250022, People's Republic of China\\
	$^{68}$ University of Manchester, Oxford Road, Manchester, M13 9PL, United Kingdom\\
	$^{69}$ University of Muenster, Wilhelm-Klemm-Strasse 9, 48149 Muenster, Germany\\
	$^{70}$ University of Oxford, Keble Road, Oxford OX13RH, United Kingdom\\
	$^{71}$ University of Science and Technology Liaoning, Anshan 114051, People's Republic of China\\
	$^{72}$ University of Science and Technology of China, Hefei 230026, People's Republic of China\\
	$^{73}$ University of South China, Hengyang 421001, People's Republic of China\\
	$^{74}$ University of the Punjab, Lahore-54590, Pakistan\\
	$^{75}$ University of Turin and INFN, (A)University of Turin, I-10125, Turin, Italy; (B)University of Eastern Piedmont, I-15121, Alessandria, Italy; (C)INFN, I-10125, Turin, Italy\\
	$^{76}$ Uppsala University, Box 516, SE-75120 Uppsala, Sweden\\
	$^{77}$ Wuhan University, Wuhan 430072, People's Republic of China\\
	$^{78}$ Yantai University, Yantai 264005, People's Republic of China\\
	$^{79}$ Yunnan University, Kunming 650500, People's Republic of China\\
	$^{80}$ Zhejiang University, Hangzhou 310027, People's Republic of China\\
	$^{81}$ Zhengzhou University, Zhengzhou 450001, People's Republic of China\\
}
	\vspace{0.2cm}
	$^{a}$ Deceased\\
	$^{b}$ Also at the Moscow Institute of Physics and Technology, Moscow 141700, Russia\\
	$^{c}$ Also at the Novosibirsk State University, Novosibirsk, 630090, Russia\\
	$^{d}$ Also at the NRC "Kurchatov Institute", PNPI, 188300, Gatchina, Russia\\
	$^{e}$ Also at Goethe University Frankfurt, 60323 Frankfurt am Main, Germany\\
	$^{f}$ Also at Key Laboratory for Particle Physics, Astrophysics and Cosmology, Ministry of Education; Shanghai Key Laboratory for Particle Physics and Cosmology; Institute of Nuclear and Particle Physics, Shanghai 200240, People's Republic of China\\
	$^{g}$ Also at Key Laboratory of Nuclear Physics and Ion-beam Application (MOE) and Institute of Modern Physics, Fudan University, Shanghai 200443, People's Republic of China\\
	$^{h}$ Also at State Key Laboratory of Nuclear Physics and Technology, Peking University, Beijing 100871, People's Republic of China\\
	$^{i}$ Also at School of Physics and Electronics, Hunan University, Changsha 410082, China\\
	$^{j}$ Also at Guangdong Provincial Key Laboratory of Nuclear Science, Institute of Quantum Matter, South China Normal University, Guangzhou 510006, China\\
	$^{k}$ Also at MOE Frontiers Science Center for Rare Isotopes, Lanzhou University, Lanzhou 730000, People's Republic of China\\
	$^{l}$ Also at Lanzhou Center for Theoretical Physics, Lanzhou University, Lanzhou 730000, People's Republic of China\\
	$^{m}$ Also at the Department of Mathematical Sciences, IBA, Karachi 75270, Pakistan\\
	$^{n}$ Also at Ecole Polytechnique Federale de Lausanne (EPFL), CH-1015 Lausanne, Switzerland\\
	$^{o}$ Also at Helmholtz Institute Mainz, Staudinger Weg 18, D-55099 Mainz, Germany\\
\end{center}
\vspace{0.4cm}
\end{small}
}

\date{\today}

\begin{abstract}
By analyzing a $\psi(3686)$ data sample containing $(107.7\pm0.6)\times10^{6}$ events 
taken with the BESIII detector at the BEPCII storage ring in 2009,
the $\chiz$ resonance parameters are precisely measured
using $\chi_{c0,c2} \to \pp/\kk$ events.
The mass of $\chiz$ is determined to be $M(\chiz)=(3415.63\pm0.07\pm0.07\pm0.07$)~MeV/$c^2$,
and its full width is $\Gamma(\chiz)=(12.52\pm0.12\pm0.13)~{\rm MeV}$, 
where the first uncertainty is statistical, the second systematic, and the third for mass comes from $\chit$ mass uncertainty.
These measurements improve the precision of $\chiz$ mass by a factor
of four and width by one order of magnitude over the previous individual measurements,
and significantly boost our knowledge about the charmonium spectrum.
Together with additional $(345.4\pm2.6)\times10^{6}$ $\psi(3686)$ data events taken in 2012, 
the decay branching fractions of $\chi_{c0,c2}\to\pp/\kk$ are measured as well,
with precision improved by a factor of three compared to previous measurements.
These $\chiz$ decay branching fractions provide important inputs for the study of glueballs.
\end{abstract}

\pacs{\color{blue}{}}

\maketitle
Charmonium, the bound state of charm and anti-charm quarks governed by the strong force, 
is analogous to the `hydrogen atom' in the study of meson spectroscopy~\cite{hydrogen,Eichten:1974af}.
Due to its heavy mass, the velocity of the charm quark is relative slow and therefore the
system can be well described by a non-relativistic potential model~\cite{potential}.
By now, the charmonium spectroscopy below the open-charm threshold is
well established~\cite{pdg},
and a precise study of them is important and necessary for a stringent
test of the theory of the strong interaction, quantum chromodynamics (QCD). 
The $\chi_{cJ}$~(spin~$J = 0,1,2)$ charmonia are $P$-wave $c\overline{c}$ states split by spin-orbit
and tensor forces. At the moment, the $\chi_{c1}$ and $\chit$ masses are precisely measured 
using a scan approach with $p\bar{p}$ annihilation by the E760~\cite{E760} and E835~\cite{E835} experiments,
and by the LHCb~\cite{LHCb} experiment in $pp$ collisions, 
whereas a precise $\chiz$ mass measurement is relatively marginal, with an uncertainty about 
five times larger~\cite{bes2,E835-chic0} than that of $\chi_{c1,c2}$. 
Improved precision on the $\chiz$ mass is important
for probing the spin structure of the strong force, 
such as the fine structure splitting $M(^3P_2)-M(^3P_0)$, 
and the singlet-triplet splitting $M(^1P_1)-M_{\rm cog}(^3P_J)$, where $M_{\rm cog}(^3P_J)$ is the center-of-gravity of the triplet.
This helps precisely determine the spin-orbit and tensor forces.

We also lack knowledge about the precise width of $\chiz$~\cite{pdg} 
compared to its $J=1,~2$ partners~\cite{E760,E835,LHCb}. 
A precise width measurement of $\chiz$
serves as an essential input for studying the
$\chiz$ decay, such as the $E1$ transition partial width, light hadron decay width, etc.
Furthermore, lattice QCD calculations show an excited scalar glueball candidate with mass
in the range of $2.8-3.7$~GeV~\cite{lattice-unquench,Gregory:2012hu,Meyer:2004gx,Chen:2005mg}, 
and the $\chiz$ might contain a gluonic admixture 
due to the presence of such a nearby glueball. In this sense, the precise $\chiz$ width provides
valuable knowledge to investigate the excited glueball spectrum, which was used to
explain the $\gamma^*\to(c\bar{c})(c\bar{c})$ cross section discrepancy~\cite{Brodsky} between 
perturbative QCD calculations~\cite{pQCD,Liu:2002wq} and the experimental measurement~\cite{belle-double}.

The decay of $\chiz$ to light hadrons proceeds predominately via two-gluon exchanges (color-singlet)~\cite{ktchao}.
The decay widths of the simplest pseudo-scalar final state $\pp$ and $\kk$ are expected to be identical within SU(3) flavor
symmetry. This simple feature also applies to other $0^{++}$ systems, such as the light glueball candidates
$f_0(1500)$ and $f_0(1710)$~\cite{pdg}. However, the couplings of $f_0(1710)$ to the $\pp$ and $\kk$ final states
differ significantly~\cite{f1710,Ablikim:2006db} due to a so-called `Chiral Suppression' effect~\cite{chiral,Chao:2007sk}.
Therefore, a precise measurement of the branching fraction (BF) $\chiz\to\pp/\kk$, 
together with its full width, provides an ideal testing ground for the study of decays of the glueball candidates.
In addition, the color-octet component also plays an important role in charmonium decays, and these precise 
BF measurements  can help to constrain the non-pertubative parameters in QCD calculations~\cite{ktchao,Bodwin:1992ye,octet} 
and to allow us to finally understand the decay dynamics of charmonium states.

In this Letter, a precise mass and width measurement of $\chiz$ is achieved via $\chiz\to\pp/\kk$ decays.
The decay BFs of $\chi_{c0,c2}\rightarrow\pi^{+}\pi^{-}/K^{+}K^{-}$ are precisely measured as well.
The analysis is performed using a $\psi(3686)$ data sample consisting of 
$(107.7\pm0.6)\times10^{6}$ events taken in 2009 and 
$(345.4\pm2.6)\times10^{6}$ events taken in 2012~\cite{BESIII:2024lks} at BESIII.
The $\chi_{c0,c2}$ states are 
produced copiously in the radiative transition $\psi(3686)\to\gamma\chi_{c0,c2}$.

The BESIII detector is described in detail elsewhere~\cite{Ablikim:2009aa}.
%
Simulated Monte Carlo~(MC) samples produced with a {\sc geant4}-based~\cite{GEANT4:2002zbu}
MC simulation software package, which includes the geometric description of the detector as well as its response,
are used to determine detection efficiencies, optimize the selection criteria, and estimate background contributions.
Signal MC samples of $\psi(3686)\rightarrow\gamma\chi_{c0,c2}\rightarrow\gamma\pi^{+}\pi^{-}/\gamma K^{+}K^{-}$
are produced. Each channel contains 400,000 signal events. 
In the simulation, the $\psi(3686)$ resonance is generated
with {\sc kkmc}~\cite{kkmc,Jadach_2000}, which includes initial-state-radiation and the beam energy spread. The decays of
$\psi(3686)\to\gamma\chi_{c0,c2}\to\gamma\pp/\gamma\kk$ are simulated
with the angular distribution taken into account using a previous multipole amplitude measurement by BESIII~\cite{BESIII:2011nst}.

To investigate the potential background, an inclusive MC sample containing the same number of $\psi(3686)$
events as data is simulated. This sample includes the production of the
$\psi(3686)$ resonance, the ISR production of the $J/\psi$, and the continuum processes incorporated in 
{\sc kkmc}~\cite{kkmc,Jadach_2000}. 
All particle decays are modeled with {\sc evtgen}~\cite{evtgen,Ping:2008zz} using BFs 
either taken from the Particle Data Group (PDG)~\cite{pdg}, when available,
or otherwise modeled with {\sc lundcharm}~\cite{ref:lundcharm,Yang:2014vra} for all remaining unknown charmonium decays.
Final state radiation~(FSR) from charged final state particles is incorporated 
using the {\sc photos} package~\cite{photos2}.
Di-muon and Bhabha MC samples, each containing one million events, are generated 
with the Babayaga generator~\cite{babayaga} for further background studies.

For the $\psi(3686)\rightarrow\gamma\chi_{c0,c2}\rightarrow\gamma\pi^{+}\pi^{-}/\gamma K^{+}K^{-}$
signal events of interest, the final states have two high-momentum charged tracks 
and an energetic radiative photon, with an energy of 261~(128)~MeV in the $\chi_{c0}~(\chi_{c2})$ case, 
due to the large mass gap between $\psi(3686)$ and $\chi_{c0,c2}$.
Charged tracks detected in the multi-layer drift chamber (MDC) 
are required to be within the polar angle range $\vert\!\cos\theta\vert<0.93$, 
where $\theta$ is defined with respect to the $z$ axis,
which is the symmetry axis of the MDC.
For each good charged track, the distance of closest approach to the interaction point (IP) 
must be less than 10\,cm
along the $z$ axis, 
and less than 1\,cm
in the transverse plane. 
A candidate event is required to have two good charged tracks with zero net charge.

Photon candidates are identified using isolated showers in the electromagnetic calorimeter (EMC).  
The deposited energy of each shower must be more than 25~MeV in the barrel region ($\vert\!\cos\theta\vert< 0.80$) 
and more than 50~MeV in the end-cap region ($0.86 <\vert\!\cos\theta\vert< 0.92$).
To exclude showers that originate from charged tracks,
the angle subtended by the EMC shower and the position of the closest charged track at the EMC
must be greater than 10 degrees as measured from the IP.
To suppress electronic noise and showers unrelated to the event, 
the difference between the EMC time and the event start time is required to be within 
[0, 700]\,ns. At least one good photon candidate is required in an event.

To improve the momentum resolution of final state particles and further suppress background,
a four-constraint (4C) kinematic fit which constrains the four-momentum of two charged tracks and a photon to the initial $\psi(3686)$
four-momentum is performed.
The two charged tracks are assumed to be either $\pp$ or $\kk$, and the corresponding 
kinematic fit chi-squares ($\chi^2_{\gamma\pp}$ and $\chi^2_{\gamma\kk}$) are obtained. An event is assigned
as $\gamma\pp$ if $\chi^2_{\gamma\pp}<\chi^2_{\gamma\kk}$, otherwise as $\gamma\kk$.
The radiative photon candidate is selected as the one that yields the smallest $\chi^2$ from the 4C kinematic fit
if there are multiple photon candidates within one event.
The $\chi^{2}$ of the kinematic fit is further required to be less than 60 
for $\gamma\pi^{+}\pi^{-}$ events and 56 for $\gamma K^{+}K^{-}$ events,
which are optimized by maximizing the Figure-of-Merit $S/\sqrt{(S + B)}$, 
where $S$ represents the number of normalized events from signal MC samples according to BFs from the PDG~\cite{pdg}, 
and $B$ is the total number of normalized background events 
estimated from the inclusive, Bhabha and di-muon MC samples.

For $\gamma\pi^{+}\pi^{-}$ events, there are radiative Bhabha and
$\psi(3686)\rightarrow(\gamma) e^{+}e^{-}$ backgrounds.  To remove
these electrons, the deposited energy of each charged track in the EMC
is required to be less than 1.34~GeV.  
Electrons passing through the EMC gap ($0.81<\left|\cos \theta\right|<0.86$), 
which has no crystals between the barrel and end-cap regions, can further survive.  
To remove these remaining electrons, the
$\mathrm{d}E/\mathrm{d}x$ of a charged particle measured by the MDC is
used. For charged tracks falling into the gap region, the
$|\chi_{{\mathrm{d}E}/{\mathrm{d}x}}(\pi)|$~\cite{Ablikim:2009aa} is
required to be less than 2, where
$\chi_{{\mathrm{d}E}/{\mathrm{d}x}}(\pi)$ is the pull value of
$\mathrm{d}E/\mathrm{d}x$ based on the pion hypothesis.  The most
significant background comes from radiative di-muon and
$\psi(3686)\to(\gamma)\MM$ events due to serious $\pi$ and $\mu$
mis-identification. These non-peaking background processes are
precisely known~\cite{pdg} and can be well simulated at BESIII.

For $\gamma K^{+}K^{-}$ events, the background level is much lower 
due to the higher kaon mass. There are some remaining backgrounds from 
radiative Bhabha, $\psi(3686)\rightarrow (\gamma) e^{+}e^{-}$, and $\psi(3686)\rightarrow \pi^{0}\pi^{0}\jpsi\to\piz\piz\EE$,
which are effectively vetoed by requiring the deposited energy of both kaons in the EMC to be less than 1.34~GeV.
Since no particle identification is applied to high momentum pions and kaons, there is cross-contamination
between $\gamma\pp$ and $\gamma\kk$ events. According to MC simulation studies, the cross-contamination
ratio is small ($<1$\%). For $\gamma\kk$ events, the background contribution from $\gamma\pp$ 
cross-contamination is fixed via MC simulation, and vice versa.

\begin{figure}[h]
 	\centering
 	\includegraphics[width=0.5\textwidth]{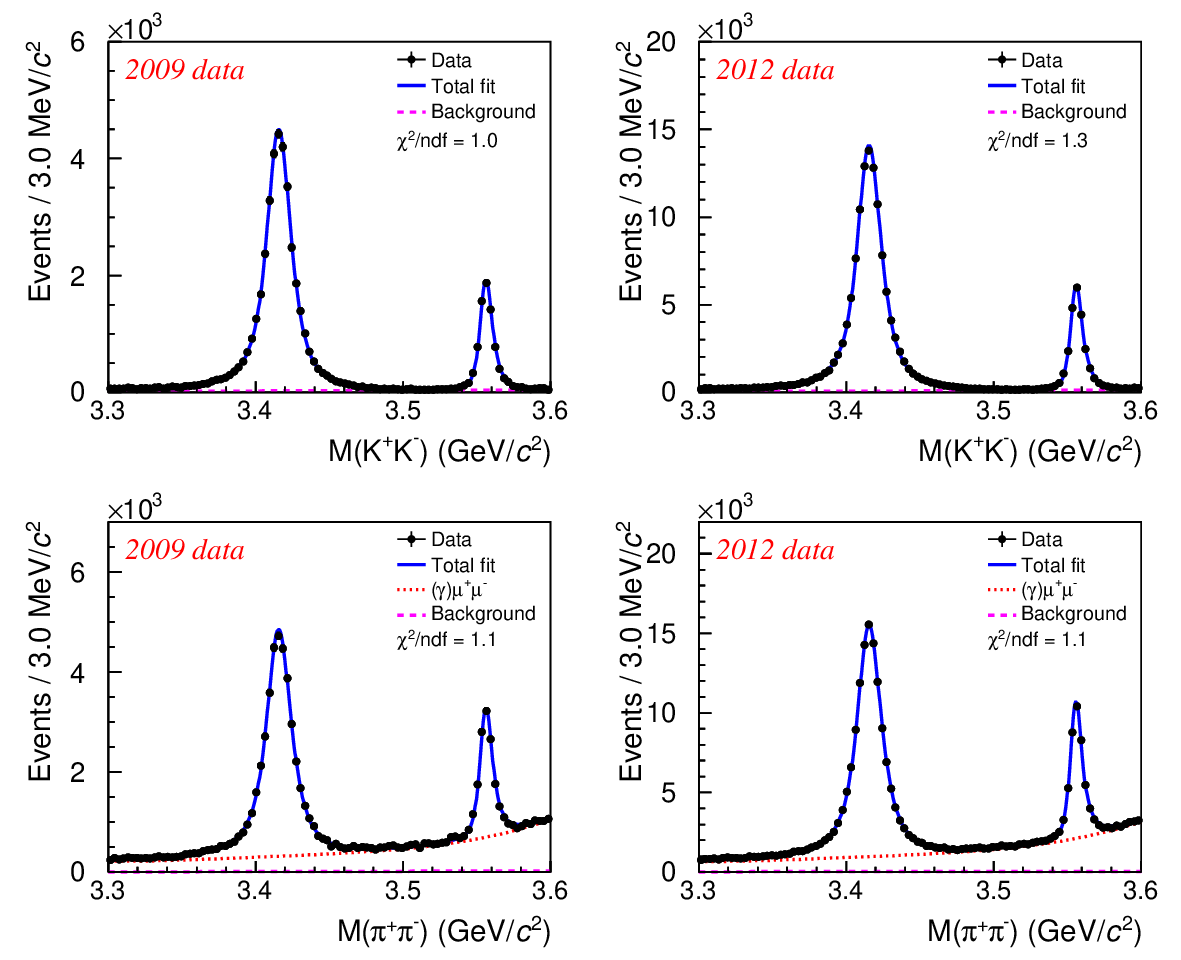}
 	\caption{The $\kk$ (top row) and $\pp$ (bottom row) invariant mass distributions from $\psi(3686)$ data taken
	in 2009 (first column) and 2012 (second column) with fit results overlaid.
	The black dots with error bars are data, the blue solid curves represent the total fit results, 
	the red dotted lines in bottom panels represent the background component from $(\gamma)\MM$, 
	and the pink dashed lines represent all other background components.}
 	\label{massandwidth}
 \end{figure}

After applying the event selection criteria mentioned above, the obtained 
$M(\pi^{+}\pi^{-})$ and $M(K^{+}K^{-})$ invariant mass distributions from data are shown in Fig.~\ref{massandwidth},
where significant numbers of $\chi_{c0,c2}$ signal events are observed. To measure the mass and width of $\chiz$, as well as the
BFs of $\chi_{c0,c2}\to\pp/\kk$, an unbinned maximum likelihood fit is performed to the $\gamma\pp$ and $\gamma\kk$ data events
simultaneously.
In the fit, the BFs of $\chi_{c0,c2}\to\pp$ and $\chi_{c0,c2}\to\kk$ are
shared as common parameters between data taken in different years.
To avoid using the absolute momentum scale, and therefore significantly improve the mass precision of $\chiz$, 
the mass split between $\chit$ and $\chiz$, $\Delta M_{20} \equiv M(\chit)-M(\chiz)$, is measured instead.
Even in this case, the quality of the 2012 data is not as good as the 2009 data 
due to the worse MDC inner chamber performance during data taking~\cite{BESIII:2017tvm}, 
which bring in considerable systematic effects from discrepancies between data and MC simulation.
Therefore, only the 2009 data are used for mass and width measurements, {\it i.e.}
$\Delta M_{20}$ and the full width $\Gamma$ of $\chiz$ are shared as common parameters
in $\gamma\pp$ and $\gamma\kk$ data events taken in 2009. 

The signal probability-density-function (PDF) is parameterized as
\begin{linenomath*}\label{pdf}
	\begin{equation}
		\begin{aligned}
            &[{\rm BW}^2(\sqrt{s}) \times \mathcal{PS}^{2}_1(\sqrt{s}) \times \mathcal{D}^{2}(\sqrt{s}) \\
			&\times \mathcal{PS}^2_2(\chi_{c0,c2}\rightarrow\pi^{+}\pi^{-}/K^{+}K^{-})\times \epsilon(\sqrt{s})]\otimes {\rm Resolution},
		\end{aligned}
	\end{equation}
\end{linenomath*}
where ${\rm BW}(\sqrt{s})$ is a relativistic Breit-Wigner (BW) amplitude defined as
\begin{linenomath*}
    \begin{equation}\label{bw}
        {\rm BW}(\sqrt{s})=\frac{1}{s-M^{2}+i M \Gamma \frac{p}{p^{\prime}}\frac{M}{\sqrt{s}}},
    \end{equation}
\end{linenomath*}
where $\sqrt{s}$ is the $\pp/\kk$ invariant mass, while $M$ and 
$\Gamma$ are the mass and constant width of $\chi_{c0,c2}$, respectively,
and $p~(p^\prime)$ is the momentum of daughter particles in the rest
frame of the mother particle with mass $\sqrt{s}~(M)$.
In the $E1$ radiative transition $\psi(3686)\to\gamma\chi_{c0,c2}$, the decay width is proportional to
a phase space factor $\mathcal{PS}_1(\sqrt{s})$~\cite{Brambilla:2010cs,Barnes:2005pb}
defined as
\begin{linenomath*}
	\begin{equation}
\mathcal{PS}_1(\sqrt{s})=(\dfrac{E_{\gamma}}{E^{0}_{\gamma}})^{3/2},
	\end{equation}
\end{linenomath*}
where $E_{\gamma}$ is the radiative photon energy and $E^{0}_{\gamma}$
corresponds to the photon energy at the $\chi_{c0,c2}$ mass $M$.
An additional damping factor 
$\mathcal{D}(\sqrt{s})=(\dfrac{(E^{0}_{\gamma})^2}{E^{0}_{\gamma} E_{\gamma}+(E^{0}_{\gamma}-E_{\gamma})^{2}})^{1/2}$~\cite{Anashin:2010dh} is also introduced to suppress the higher energy tail.
$\mathcal{PS}_2(\chi_{c0,c2}\rightarrow\pi^{+}\pi^{-}/K^{+}K^{-})$ is the phase space factor of $\chi_{c0,c2}$
decaying into $\pp/\kk$~\cite{pdg}.
The signal PDF is corrected by a mass-dependent efficiency curve $\epsilon(\sqrt{s})$, 
which is obtained by MC simulation studies and varies within 1\% for the interested mass region, thus parameterized as a linear function. 
The $\chi_{c0,c2}$ resonance line shapes are further convolved with
Gaussian functions to account for the detector resolution.
In the fit, the $\chit$ width is fixed to the known value~\cite{pdg}, and the resolution difference between data and MC simulation 
measured from $\chit$ data events is used to calibrate that for $\chiz$.

The shapes of dominant background sources are described by polynomial
functions. For $\gamma\kk$ events, the background is low and
parameterized as a free second-order polynomial. The small
cross-contamination contribution from $\gamma\pp$ events is fixed via
MC simulation.  For $\gamma \pi^{+} \pi^{-}$ events, the dominating
$(\gamma)\MM$ background is represented by a fourth-order polynomial,
which is fixed according to normalized MC events for
$\psi(3686)\to(\gamma)\MM$ and di-muon processes.  Other extra
backgrounds with a small contribution ($\sim 3\%$ of the total
background) are parameterized as a free second-order polynomial.

Figure~\ref{massandwidth} shows the fit results. The mass split between $\chit$ and $\chiz$
is measured to be $\Delta M_{20}\equiv M(\chi_{c 2})-M(\chi_{c 0}) = (140.54\pm0.07$)~MeV/$c^2$.
Taking the world average mass $M(\chit)=(3556.17\pm0.07)$~MeV/$c^2$~\cite{pdg} as input, 
the $\chiz$ mass is calculated to be $M(\chiz)=(3415.63\pm0.07$)~MeV/$c^2$.
The width of $\chi_{c0}$ is measured to be $\Gamma(\chi_{c0}) = (12.52\pm 0.12) ~\rm{MeV}$.
The uncertainties here are statistical only.

The systematic uncertainties in the mass split and $\chiz$ width measurements 
mainly come from 
the signal and background PDF shapes as well as the damping factor, 
momentum-dependent scale, and possible interference effect.
In the nominal fit model, a BW function with a mass-dependent width, {\it i.e.} Eq.~\ref{bw}, 
is used to describe the $\chi_{c0,c2}$ state.
An alternative shape with a constant width BW function is investigated. 
The difference with respect to the nominal measurement
for $\Delta M_{20}$ is 0.04~MeV/$c^2$ and for the $\chiz$ full width $\Gamma(\chiz)$ is 0.08~MeV, 
which are taken as systematic uncertainties.
A damping factor also appears in the signal PDF. We take another form
$\mathcal{D}(\sqrt{s})=e^{-{E_{\gamma}^2}/({8\beta^2})}$, where $\beta=65.0\pm2.5~\rm{MeV}$
from the CLEO-c Collaboration~\cite{CLEO:2008pln}.
The difference on the measurements of $\Delta M_{20}$
and $\Gamma(\chiz)$ is 0.01~MeV/$c^2$ and 0.01~MeV, respectively.

The uncertainty due to background in the fit is studied in several aspects.
The free background components are described by second-order polynomials for both $\gamma\pp$ and $\gamma\kk$ events.
Changing this component to a third-order polynomial either for $\gamma\pp$ or $\gamma\kk$ events 
yields differences on the measurements, which are 0.02~MeV/$c^2$ for $\Delta M_{20}$ and 0.04~MeV for $\Gamma(\chiz)$.
Interference between $\chi_{c0,c2}$ resonances and continuum is observed in low-mass regions~\cite{Dobbs:2015dwa},
and we account for this by adding the continuum component coherently to the signal PDF.
The resulting difference for $\Delta M_{20}$ is 0.01~MeV/$c^2$ and $\Gamma(\chiz)$ is 0.08~MeV.
For $\gamma\pp$ events, the dominant background comes from $(\gamma)\MM$.
Varying the output cross section from Babayaga~\cite{babayaga} by $\pm1\sigma$ 
for MC simulated background normalization
yields a difference of 0.02~MeV/$c^2$ for $\Delta M_{20}$ and 0.02~MeV for $\Gamma(\chiz)$.
The shape parameters of this background component are also varied within $\pm1\sigma$,
which yields a difference of 0.01~MeV/$c^2$ for $\Delta M_{20}$ and 0.02~MeV for $\Gamma(\chiz)$.
The momentum-dependent scale is simulated by MC. The systematic uncertainty due
to the difference between data and MC is studied by correcting the MC distribution
of final-state particle momenta to data with $\chit$ events,
which yields a variation of 0.04~MeV/$c^2$ for $\Delta M_{20}$ and 0.01~MeV for $\Gamma(\chiz)$.

Table~\ref{table massandwidth} summarizes all these sources and
their corresponding contributions.
Assuming all the sources are independent, the total systematic uncertainty is obtained
by adding each individual source in quadrature, which is 0.07~MeV/$c^2$ for $\Delta M_{20}$ 
and 0.13~MeV for $\Gamma(\chiz)$.
\begin{table}[h]
 	\centering
 	\caption{Sources of systematic uncertainties and their contributions to $\Delta M_{20}$ and $\Gamma(\chiz)$ measurements.}\label{table massandwidth}
 	\begin{tabular}{ccccc}
 	\hline\hline
        &\multirow{2}{*}{Source}& $\Delta M_{20}$ & $\Gamma(\chiz)$  \\ 
        & &($\rm{MeV/c^2}$) &($\rm{MeV}$) \\
	\hline
         &Signal shape& $0.04$ & $0.08$  \\
         &Damping factor & $0.01$ & $0.01$  \\
         &Polynomial background & $0.02$ & $0.04$  \\
         &Interference effect & 0.01 & 0.08 \\
         &Number of $(\gamma)\MM$ background & $0.02$ & $0.02$  \\
         &Shape of $(\gamma)\MM$ background & $0.01$ & $0.02$  \\
         &Momentum dependent scale & $0.04$ & $0.01$  \\
         \hline
         &Total& $0.07$ & $0.13$    \\
 	\hline\hline
 	\end{tabular}
 \end{table}

The BF of $\chi_{c0,c2}\to\pp/\kk$ is calculated as 
\begin{equation}
\mathcal{B}=\frac{N_{\rm sig}}{N_{\rm tot}[\psi(3686)]\epsilon\mathcal{B}[\psi(3686)\to\gamma\chi_{c0,c2}]},
\end{equation}
where $N_{\rm sig}$ is the number of observed $\chi_{c0,c2}$ signal events from the fit,
$N_{\rm tot}[\psi(3686)]$ is the total number of collected $\psi(3686)$ events~\cite{BESIII:2024lks}, $\epsilon$ is the detection efficiency of $\gamma\pp/\gamma\kk$ events,
and $\mathcal{B}[\psi(3686)\to\gamma\chi_{c0,c2}]$ is the BF of the $E1$ radiative transition~\cite{pdg}.
In practice, a simultaneous fit is performed and the BFs are shared between different data sets.
The BFs are measured to be
$\mathcal{B}(\chi_{c0} \rightarrow K^{+} K^{-})=(6.36 \pm 0.02) \times 10^{-3}$,
$\mathcal{B}(\chi_{c0} \rightarrow \pi^{+} \pi^{-})=(6.06 \pm 0.02) \times 10^{-3}$, 
$\mathcal{B}(\chi_{c2} \rightarrow K^{+} K^{-})=(1.22 \pm 0.01) \times 10^{-3}$,  
and $\mathcal{B}(\chi_{c2} \rightarrow \pi^{+} \pi^{-})=(1.61 \pm 0.01) \times 10^{-3}$,
respectively, where all the uncertainties are statistical only.
Table~\ref{BF} lists the detection efficiencies for each data set.

\begin{table}[h]
	\centering 
	\caption{The detection efficiencies for $\gamma\pp$ and $\gamma\kk$ data events taken in different years.}\label{BF}	
	\begin{tabular}{ccccccc}			
		\hline\hline			
		& & $\varepsilon(\gamma\pi^{+}\pi^{-})$ &$\varepsilon(\gamma K^{+}K^{-})$     \\			
		\hline			
		&2009 $\chi_{c0}$& $0.644\pm0.001$& $0.589\pm0.001$  \\            
		&2009 $\chi_{c2}$& $0.667\pm0.001$& $0.624\pm0.001$  \\			
		&2012 $\chi_{c0}$& $0.634\pm0.001$& $0.577\pm0.001$  \\   		
		&2012 $\chi_{c2}$& $0.658\pm0.001$& $0.613\pm0.001$  \\			
		\hline\hline		
\end{tabular}\end{table}

The systematic uncertainties in the BF measurements
mainly come from the total number of $\psi(3686)$ events, detection efficiencies, signal and background PDF shapes,
possible interference effect, fit range, 4C kinematic fit, and intermediate BFs.

The total number of $\psi(3686)$ events is measured by counting inclusive hadronic events, 
with an uncertainty of $0.7\%$~\cite{BESIII:2024lks}.
Detection efficiencies include tracking, photon detection, decay angular distributions, and selection criteria.
The uncertainty of photon detection is measured to be 0.5\% per photon
by studying the $e^+e^- \rightarrow \gamma \mu^+\mu^-$ process
and 0.5\% is assigned for this analysis.
Pion and kaon tracks have a momentum above 1~GeV. The tracking efficiency of such tracks is above $99\%$ at BESIII,
by studying $\EE\to\pp\pp$ and $\jpsi\to K^*K$ events. According to the two-dimensional (2D, $p_t$ vs. $\cos\theta$)
tracking efficiencies of data and MC simulation, the uncertainty in each bin is obtained. The final uncertainty 
due to tracking is estimated as a weighted average over 2D bins.

Decay angular distributions also affect the detection efficiencies.
Considering the helicity amplitudes and correlation coefficients measured by BESIII~\cite{BESIII:2011nst},  
100 sets of two correlated helicity amplitude parameters are obtained via a 2D sampling. 
The resulting changes in detection efficiencies, and therefore the corresponding changes of BFs 
from the simultaneous fit, are taken as systematic uncertainties.
The uncertainties due to the pion/kaon EMC deposit energy requirement and fit range are studied
using a Barlow test~\cite{barlow},
by changing a series of EMC deposit energies and fit ranges.
All the alternative choices give results within 1$\sigma$ of the nominal result, 
which means these two sources can be ignored. 

For the 4C kinematic fit, a correction of the pull distribution of charged track parameters is applied to
the MC simulation~\cite{BESIII:2012mpj}. The uncertainties are taken as half of the efficiency 
difference with or without helix parameter corrections.
The uncertainties on the BFs for $\psi(3686)\rightarrow\gamma\chi_{c0}$ ($2.1\%$) and $\psi(3686)\rightarrow\gamma\chi_{c2}$ ($2.1\%$) are taken from the PDG~\cite{pdg}. 
The uncertainties due to the signal shape, damping factor and background are studied using the same method as
the mass split and width measurement.
The cross-contaminations between $\gamma\pp$ and $\gamma\kk$ events are studied via
MC simulations, with input BFs updated by this measurement. The fluctuations due to BFs
are taken as systematic uncertainties.

Table~\ref{table BF} summarizes all these sources and their contributions to the BF measurements.
Assuming all the sources are independent, the total systematic uncertainties are obtained 
by adding them in quadrature.
The largest systematic uncertainty comes from the intermediate BFs,
and all other sources contribute about 1.2\%.

\begin{table}[h]
	\centering	
	\caption{ Sources of systematic uncertainties and their contributions (in $\%$) to BF measurements.
	``..." means a non-applicable option.}\label{table BF}
	\begin{tabular}{ccccccc}			
		\hline\hline			
		&\multirow{2}{*}{Source} & \multicolumn{2}{c}{$\gamma\pi^{+}\pi^{-}$} &\multicolumn{2}{c}{$\gamma K^{+}K^{-}$}  \\            
		&  & $\chi_{c0}$ & $\chi_{c2}$  & $\chi_{c0}$ &$\chi_{c2}$  \\			
		\hline			
		&$N_{\rm tot}$[$\psi(2S)$] & $0.7$ & $0.7$ & $0.7$ & $0.7$   \\ 			
		&Tracking & $0.1$ & $0.1$  & $0.2$ & $0.1$  \\			
		&Photon & $0.5$ & $0.5$ & $0.5$ & $0.5$ \\			
		&Polynomial background & $0.6$ & $0.2$ & $0.1$ & $0.1$    \\      
		&Interference effect & 0.4 & 0.0 & 0.4 & 0.0 \\      
		&Number of $(\gamma)\mu^{+}\mu^{-}$ background  & $0.1$ & $0.4$ & $0.1$ & $0.1$    \\	
		&Shape of $(\gamma)\mu^{+}\mu^{-}$ background & $0.2$ & $0.1$ & $0.1$ & $0.1$    \\		
		&Cross contamination  & 0.0 & 0.0 & 0.0 & $0.1$ \\			
		&Damping factor  & $0.1$ & $0.2$& $0.1$ & $0.1$    \\			
		&Signal shape& $0.1$ & $0.2$ & $0.1$ & $0.1$  \\			
		&Angular distribution& ... & $0.1$& ... & 0.0 \\   	
		&4C kinematic fit& $0.3$ & $0.2$ & $0.7$ & $0.8$ \\		
		\hline   			
		&Sum& $1.2$ & $1.1$& $1.2$ & $1.2$\\			
		\hline			
		&Intermediate BF& $2.1$ & $2.1$  & $2.1$ & $2.1$  \\			
		\hline			
		&Total & $2.4$ & $2.4$& $2.4$ & $2.4$ \\			
		\hline\hline		
	\end{tabular}	
\end{table}

\begin{figure}[h]
 	\centering
 	\includegraphics[width=0.45\textwidth]{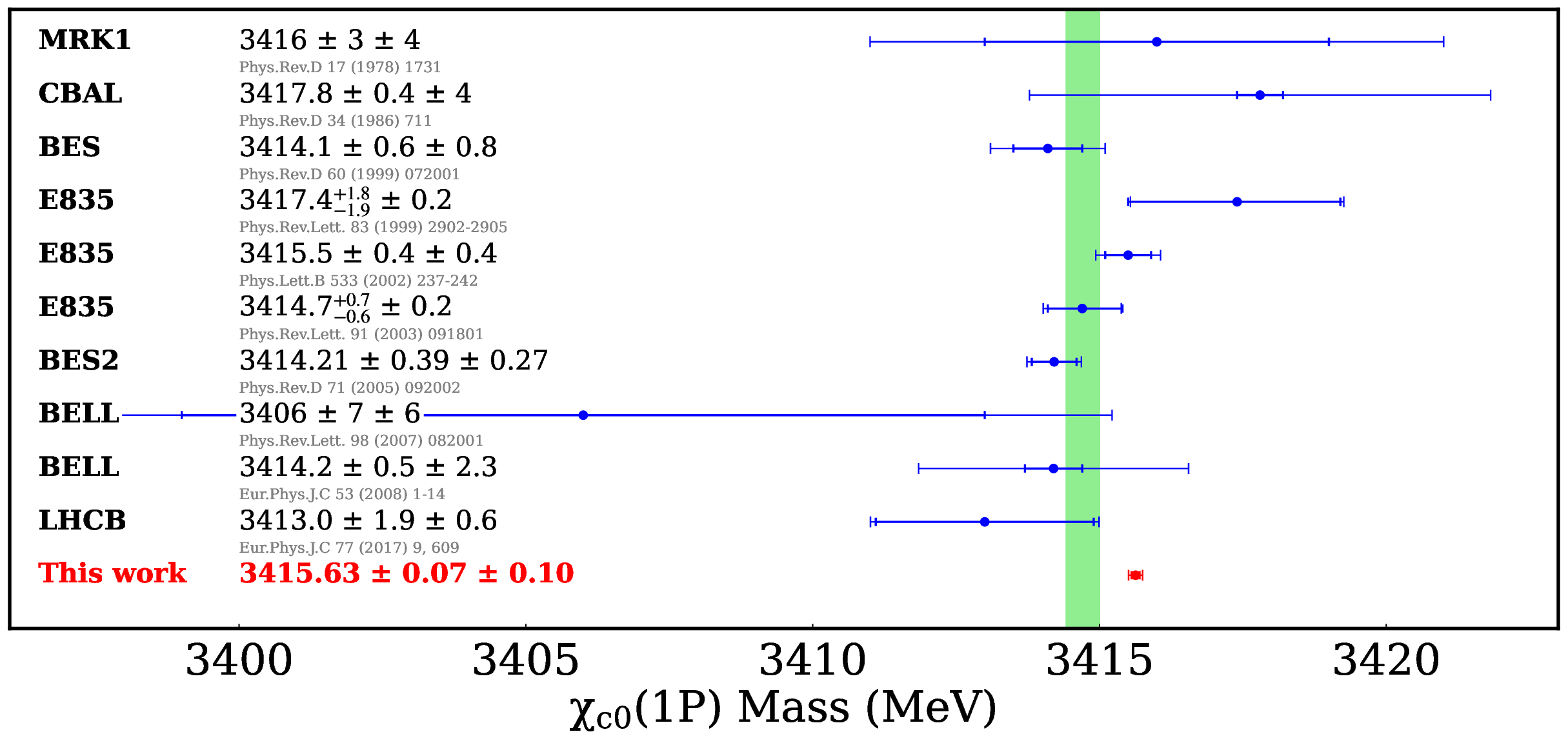}
 	\includegraphics[width=0.45\textwidth]{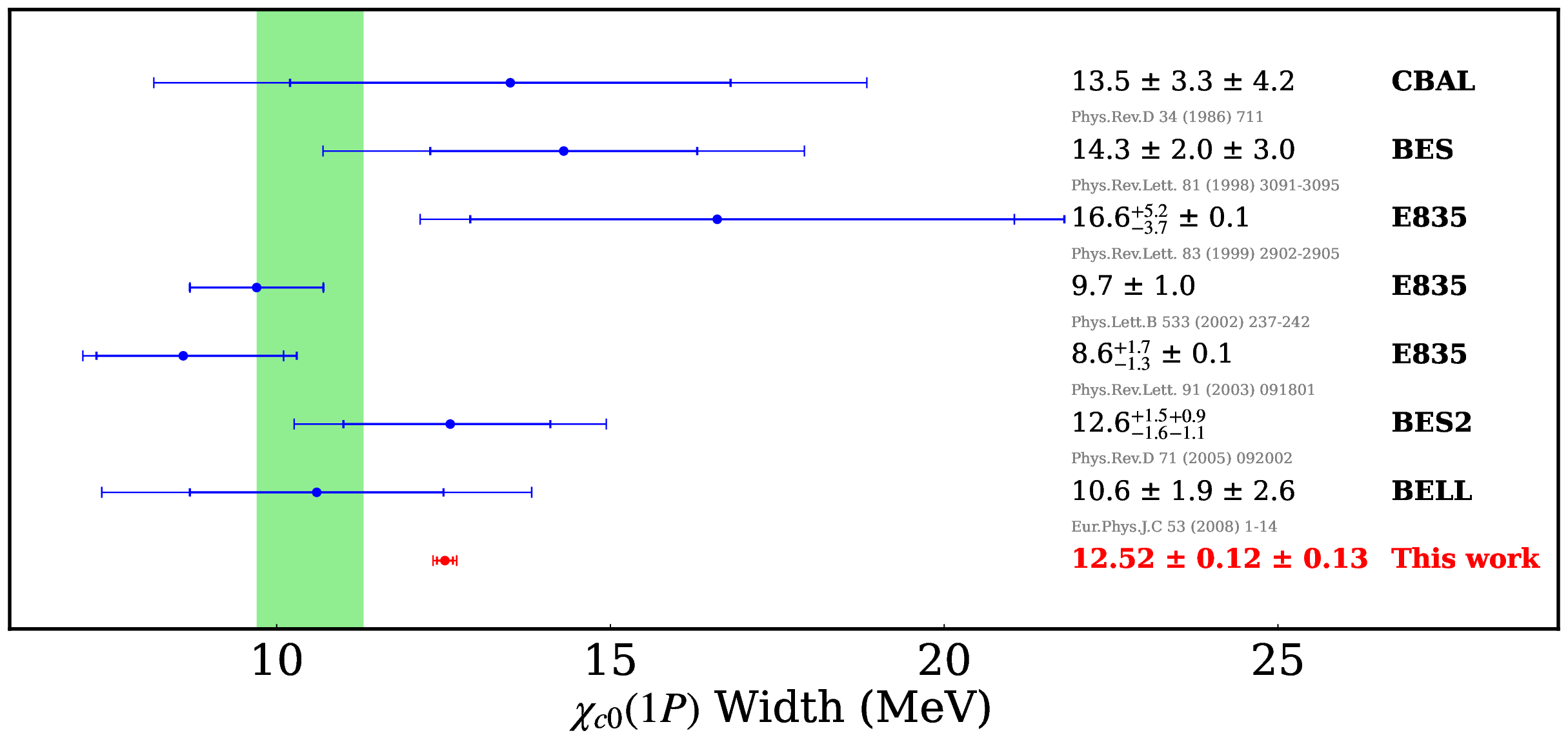}
 	\caption{Comparison of the $\chiz$ resonance parameters measurements. 
	The light green bands represent the PDG world average~\cite{pdg}.}
 	\label{compare}
 \end{figure}

In summary, a precise measurement of the $\chiz$ resonance parameters
and the $\chi_{c0,c2}\to\pp/\kk$ decay BFs is performed at BESIII
by analyzing $\psi(3686)$ data events.
The mass split between $\chi_{c2}$ and $\chi_{c0}$ is measured to be $\Delta M_{20}=(140.54\pm0.07\pm0.07)~{\rm MeV}/c^2$.
Taking the $\chit$ world average mass as input, the $\chiz$ mass is determined to be
$M(\chiz)=(3415.63\pm0.07\pm0.07\pm0.07$)~MeV/$c^2$, 
where the first uncertainty is statistical, the second systematic, and the third from the $\chit$ mass uncertainty.
This is the most precise $\chiz$ mass measurement to date. 
It is a bit higher ($1.42\pm0.49$~MeV/$c^2$) than the BES measurement~\cite{bes2},
while it agrees well with the E835 measurement~\cite{E835-chic0}.
Our measurement improves the precision of the $\chiz$ mass over previous individual
measurements by a factor of four~\cite{bes2,E835-chic0}.
The full width of $\chi_{c0}$ is measured to be $\Gamma(\chiz)=(12.52\pm0.12\pm0.13)~{\rm MeV}$.
It agrees with the BES measurement quite well within $1\sigma$~\cite{bes2}
and improves the precision by an order of magnitude.
These measurements have a big impact on our knowledge about the $\chiz$ resonance parameters~\cite{pdg},
and are expected to play a crucial role in our understanding of the strong force governing the $c\bar{c}$ 
system~\cite{Lucha:1991vn,Barnes:2005pb}.
%
A comparison of our results with other measurements as well as the PDG
world average is shown in Fig.~\ref{compare}.

The decay BFs are measured to be 
\begin{eqnarray}\notag
	\mathcal{B}(\chiz\to\kk)=(6.36\pm0.02\pm0.08\pm0.13)\times10^{-3}, \\\notag
	\mathcal{B}(\chiz\to\pp)=(6.06\pm0.02\pm0.07\pm0.13)\times10^{-3}, \\\notag
	\mathcal{B}(\chit\to\kk)=(1.22\pm0.01\pm0.02\pm0.03)\times10^{-3}, \\\notag
	\mathcal{B}(\chit\to\pp)=(1.61\pm0.01\pm0.02\pm0.04)\times10^{-3}, \\\notag
\end{eqnarray}
where the first uncertainties are statistical, the second systematic, 
and the third come from $\mathcal{B}[\psi(3686)\rightarrow \gamma\chi_{c0,c2}]$~\cite{pdg}.
These measurements agree with the CLEO-c result within $1\sigma$~\cite{cleo-BF},
and improve the precision by more than threefold.
Interestingly, the ratio of BFs
$\mathcal{R}=\frac{\mathcal{B}(\chiz\to\pp)}{\mathcal{B}(\chiz\to\kk)}=(0.95\pm0.01\pm0.01)$
demonstrates no obvious `Chiral Suppression' is observed for this $0^{++}$ system,
which supports the theoretical analysis in Ref.~\cite{Chao:2005si,Zhang:2005mu}.
This behavior is also quite different from the case of the $f_0(1710)$~\cite{f1710,Ablikim:2006db},
therefore challenging the scalar glueball interpretation of the $f_0(1710)$ state~\cite{chiral}.
It implies $f_0(1710)$ might contain a substantial $s\bar{s}$ component,
which aligns with the model calculations in Ref.~\cite{Close:2005vf,Narison:2005wc}.
Combining our precisely measured $\mathcal{R}$ value, which constrains the
glueball fraction in $f_0(1710)$, with future theoretical improvements
will advance our understanding of QCD.

The BESIII Collaboration thanks the staff of BEPCII and the IHEP computing center for their strong support. We are also grateful to Prof. Chen Ying for the inspiring discussion on glueballs. This work is supported in part by National Key R\&D Program of China under Contracts Nos. 2020YFA0406300, 2020YFA0406400, 2023YFA1606000; National Natural Science Foundation of China (NSFC) under Contracts Nos. 11635010, 11735014, 11935015, 11935016, 11935018, 12025502, 12035009, 12035013, 12061131003, 12192260, 12192261, 12192262, 12192263, 12192264, 12192265, 12221005, 12225509, 12235017, 12361141819; the Chinese Academy of Sciences (CAS) Large-Scale Scientific Facility Program; the CAS Center for Excellence in Particle Physics (CCEPP); Joint Large-Scale Scientific Facility Funds of the NSFC and CAS under Contract No. U1832207; 100 Talents Program of CAS; 
Project
No. ZR2022JQ02, ZR2024QA151 supported by Shandong Provincial
Natural Science Foundation; supported by the China
Postdoctoral Science Foundation under Grant
No. 2023M742100; 
The Institute of Nuclear and Particle Physics (INPAC) and Shanghai Key Laboratory for Particle Physics and Cosmology; German Research Foundation DFG under Contracts Nos. FOR5327, GRK 2149; Istituto Nazionale di Fisica Nucleare, Italy; Knut and Alice Wallenberg Foundation under Contracts Nos. 2021.0174, 2021.0299; Ministry of Development of Turkey under Contract No. DPT2006K-120470; National Research Foundation of Korea under Contract No. NRF-2022R1A2C1092335; National Science and Technology fund of Mongolia; National Science Research and Innovation Fund (NSRF) via the Program Management Unit for Human Resources \& Institutional Development, Research and Innovation of Thailand under Contracts Nos. B16F640076, B50G670107; Polish National Science Centre under Contract No. 2019/35/O/ST2/02907; Swedish Research Council under Contract No. 2019.04595; The Swedish Foundation for International Cooperation in Research and Higher Education under Contract No. CH2018-7756; U. S. Department of Energy under Contract No. DE-FG02-05ER41374.

\bibliography{reference}

\end{document}